\documentclass[pdflatex,sn-mathphys-num]{sn-jnl}

\usepackage{graphicx}%
\usepackage{multirow}%
\usepackage{amsmath,amssymb,amsfonts}%
\usepackage{amsthm}%
\usepackage{mathrsfs}%
\usepackage[title]{appendix}%
\usepackage{xcolor}%
\usepackage{textcomp}%
\usepackage{manyfoot}%
\usepackage{booktabs}%
\usepackage{algorithm}%
\usepackage{algorithmicx}%
\usepackage{algpseudocode}%
\usepackage{listings}%

\usepackage{etoolbox}
\apptocmd{\thebibliography}{\setlength{\itemsep}{0pt}}{}{}

\theoremstyle{thmstyleone}%
\theoremstyle{thmstyletwo}%

\theoremstyle{thmstylethree}%

\begin{document}

\title[Simulating User Watch-Time to Investigate Bias in YouTube Shorts Recommendations]{Simulating User Watch-Time to Investigate Bias in YouTube Shorts Recommendations}


\author[1]{\fnm{Selimhan} \sur{Dagtas}}\email{sedagtas@ualr.edu}
\author[1]{\fnm{Mert Can} \sur{Cakmak}}\email{mccakmak@ualr.edu}
\author[1,2]{\fnm{Nitin} \sur{Agarwal}}\email{nxagarwal@ualr.edu}

\affil[1]{COSMOS Research Center, University of Arkansas, Little Rock, USA}

\affil[2]{ICSI, University of California, Berkeley, USA}

\abstract{

Short-form video platforms such as YouTube Shorts increasingly shape how information is consumed, yet the effects of engagement-driven algorithms on content exposure remain poorly understood. This study investigates how different viewing behaviors, including fast scrolling or skipping, influence the relevance and topical continuity of recommended videos. Using a dataset of over 404,000 videos, we simulate viewer interactions across both broader geopolitical themes and more narrowly focused conflicts, including topics related to Russia, China, the Russia–Ukraine War, and the South China Sea dispute. We assess how relevance shifts across recommendation chains under varying watch-time conditions, using GPT-4o to evaluate semantic alignment between videos. Our analysis reveals patterns of amplification, drift, and topic generalization, with significant implications for content diversity and platform accountability. By bridging perspectives from computer science, media studies, and political communication, this work contributes a multidisciplinary understanding of how engagement cues influence algorithmic pathways in short-form content ecosystems.

}

\keywords{YouTube Shorts, Algorithmic Bias, Recommender Systems, Generative AI, Content Relevance, Watch-Time}

\maketitle

\section{Introduction}\label{sec1}

Understanding how recommendation systems influence content exposure is a critical concern across disciplines, including computer science, political communication, and media studies. As digital platforms increasingly mediate access to information, even subtle algorithmic design choices can shape public discourse, reinforce societal narratives, and affect user perception. YouTube Shorts, with its rapid, passive engagement model and global reach, presents a unique environment to study these dynamics. Investigating how short-form recommendation systems respond to user interactions is essential for improving algorithmic fairness and transparency, and for informing broader debates on media influence and information integrity.

Short-form video platforms have transformed digital content consumption, with YouTube Shorts emerging as a major force since its 2021 launch. The platform now attracts over 1.5 billion monthly users \cite{chan2022shorts}, and its recommendation engine heavily influences what viewers encounter. While prior research has examined algorithmic behavior in long-form content, short-form systems remain less understood. Recent studies highlight potential bias in Shorts recommendations, including content drift and visual manipulation \cite{cakmak2025unpacking, cakmak2024unveiling}. Unlike long videos, Shorts rely almost exclusively on lightweight engagement signals such as swiping and watch duration, demanding new frameworks for assessing algorithmic influence.

This study investigates how simulated watch-time behavior affects the topical relevance of content recommended by YouTube Shorts. We pose the following research questions:

\begin{itemize}
    \item \textbf{RQ1:} How does user watch-time behavior influence the topical relevance of videos recommended by the YouTube Shorts algorithm over time?
    \item \textbf{RQ2:} Does the use of interest-based watch times amplify recommendation relevance compared to uniform short watch durations?
    \item \textbf{RQ3:} To what extent does the YouTube Shorts algorithm retain topical specificity in recommendations for broader versus more specific topics?
\end{itemize}

To explore these questions, we conducted an empirical analysis using four thematic datasets: \textit{Broader Russia}, \textit{Russia–Ukraine War}, \textit{Broader China}, and \textit{South China Sea Dispute}. These topics were selected for their geopolitical importance and visibility in online discourse. The Russia–Ukraine conflict represents a major war with global implications \cite{masters2023ukraine}, while the South China Sea dispute involves contested maritime claims central to regional security and international trade \cite{bbc2023southchinasea}. Together, these cases offer meaningful testbeds for examining how recommendation systems manage both broad and specific narratives in politically sensitive domains.

\section{Literature Review}

Understanding YouTube’s recommendation dynamics, especially in short-form formats like Shorts, requires a multidisciplinary lens across computer science, communication, and social computing. Watch-time has become a dominant signal in YouTube’s algorithm, replacing earlier click-based models. The study \cite{covington2016deep} describe the platform’s two-stage recommendation pipeline optimized for user retention. However, engagement-centric systems raise concerns about feedback loops and algorithmic bias, particularly in politically sensitive contexts where exposure is shaped by personalization \cite{ribeiro2020auditing}.

To address such biases, various strategies have been proposed, including diversity-promoting re-ranking \cite{liu2019personalized}, exposure-aware training objectives \cite{mehrotra2018towards}, and fairness-aware ranking models \cite{singh2018fairness}. These methods aim to balance relevance with diversity, though their use in short-form video settings remains limited.

Network-based approaches provide further insight into how content clusters evolve in recommendation systems. Community detection methods such as Louvain \cite{blondel2008fast} reveal structural polarization, while recent work emphasizes how recommender systems shape content propagation and exposure trajectories \cite{ng2023exploring}. Although prior studies have explored recommendation behavior and bias, few have simulated how varying watch-time alone influences algorithmic responses. This study fills that gap by modeling fast-skipping versus longer viewing in YouTube Shorts to reveal how attention patterns affect topical relevance and recommendation drift.

\section{Methodology}

In this section, we present the simulation framework used to examine how varying watch-time behaviors influence the relevance of YouTube Shorts recommendations.

\subsection{Data Collection}

To examine how YouTube Shorts responds to different watch-time behaviors, we collected recommendation data across four themes: Russia–Ukraine War, Broader Russia, South China Sea Conflict, and Broader China. This design contrasts specific, fast-evolving geopolitical conflicts with broader, more stable narratives. Keywords were curated from news sources, policy reports, and academic literature, then refined using language models. For example, Russia–Ukraine terms focused on military and humanitarian developments \cite{guardian2023warcrimes}, while South China Sea keywords emphasized maritime disputes \cite{wilson2023coc}. Broader themes reflected topics such as Soviet nostalgia \cite{cepa2022sovietnostalgia} and China's global posture \cite{csis2024navalbuildup}. These keywords guided both search and analysis, allowing systematic comparison of algorithmic responses across topic types. Representative keywords are shown in Table~\ref{tab:keywords}. All data were collected from publicly available metadata using non-authenticated browsing sessions.\footnote{We scraped only publicly visible video titles and recommendations without logging in or bypassing access restrictions. Queries were rate-limited, and no private data were accessed, aligning with common academic practices for ethical web data collection.}

\begin{table}[h]
\caption{Summary of Keywords Used for Data Collection. Representative keywords are shown for each dataset.}
\label{tab:keywords}
\begin{tabular*}{\textwidth}{@{\extracolsep\fill}lp{0.8\textwidth}}
\toprule
\textbf{Dataset} & \textbf{Keywords (Selected)} \\
\midrule
Russia-Ukraine & Ukraine frontline update, Russia missile attack, NATO and Ukraine, Ukraine war explained, Ukraine refugee crisis, Russian war crimes Ukraine, Crimea missile strike, Ukraine reconstruction plan. \\
Broader Russia & Russia global influence, Russia-China alliance, Russia BRICS expansion, Russkiy Mir ideology, Russia propaganda abroad, Soviet Union expansion, Russia soft power diplomacy, Russia foreign policy strategy. \\
South China Sea & South China Sea dispute, China nine-dash line, Freedom of navigation operations, Scarborough Shoal standoff, China maritime militia, ASEAN South China Sea talks, China island militarization, US Navy South China Sea. \\
Broader China & China global expansion, Belt and Road Initiative, China military buildup, China debt trap diplomacy, Confucius Institutes controversy, China cyber warfare capabilities, China United Nations influence, China global leadership ambitions. \\
\botrule
\end{tabular*}
\end{table}

\textbf{YouTube Shorts and Recommendation Collection} — Since the YouTube Data API v3 does not support Shorts, we used APIFY’s YouTube Scraper \cite{streamers_2024_youtube} to collect 1,000 seed videos per topic (4,000 total). To simulate recommendation dynamics, we developed a custom Selenium-based scraper, as no existing tool captures Shorts recommendations directly. Using a headless Chrome driver, we automated scrolling behavior in isolated incognito sessions with cleared cookies to reduce personalization. Each seed was tested under two viewing conditions: (1) minimal viewing (3 seconds) and (2) interest-based viewing (3, 15, or 60 seconds), with durations dynamically assigned based on title relevance via a generative AI model (see Section~\ref{Relevancy-methodology}). Each session extended to a depth of 50, resulting in 400,000 recommended videos and 404,000 Shorts overall.

These datasets were selected to reflect a balance between broad geopolitical narratives (e.g., China's rise, Russia’s global influence) and specific conflicts (e.g., Russia–Ukraine War, South China Sea Dispute), offering a comparative lens across general and granular topics. The chosen themes are frequently present in trending YouTube Shorts and were validated during keyword pre-sampling, making them highly relevant to platform users and algorithmic behavior.

\subsection{Relevancy Measurement} \label{Relevancy-methodology}

We emphasize that all recommendations were collected directly from YouTube using a custom browser automation framework; the LLM was not used to generate recommendations, but only to score their topical relevance. To estimate the topical relevance of recommended YouTube Shorts, we employed generative AI models to score video titles during the scraping process. This automated scoring guided simulated watch behavior in the recommendation collection. Each model received the following prompt: \textit{You are an expert assistant specializing in assessing how relevant a YouTube video title is to topics surrounding \textbf{[X]}. For a given YouTube video title, assign a relevance score: 2 for highly relevant, 1 for somewhat relevant, and 0 for irrelevant.} The placeholder \textbf{[X]} was dynamically filled with a short description of the dataset theme (e.g., South China Sea maritime conflict, Russia–Ukraine war, China’s global expansion, or Russia’s foreign policy strategy), as defined earlier.

\textbf{Model Selection and Validation} — To evaluate model performance for this task, we tested LLaMA 3, GPT-4o, and Gemini 1.5 on two public QA/relevance datasets: MS MARCO~\cite{bajaj_marco} and WikiQA~\cite{yang_wikiqa}. As shown in Table~\ref{tab:validation}, GPT-4o achieved the highest average accuracy and was used in our analysis.

This approach enables scalable relevance scoring across 400,000 videos. QA datasets were used for validation, as they reflect the core task of assessing title-topic alignment. Given the limited metadata in Shorts, LLM-based scoring is a practical solution. To build trust, we manually validated 100 samples, with over 85\% agreement, confirming GPT-4o's consistency.

\begin{table}[h]
\caption{Validation accuracy (\%) on publicly available QA/relevance datasets used to benchmark the scoring behavior of large language models.}
\label{tab:validation}
\centering
\begin{tabular}{lccc}
\toprule
\textbf{Dataset} & \textbf{LLaMA 3} & \textbf{GPT-4o} & \textbf{Gemini 1.5} \\
\midrule
MS MARCO~\cite{bajaj_marco} & 71.80 & 80.46 & 69.84 \\
WikiQA~\cite{yang_wikiqa}   & 77.50 & 77.00 & 80.60 \\
\textbf{Average}            & 74.65 & \textbf{78.73} & 75.22 \\
\botrule
\end{tabular}
\end{table}

\section{Results}

\begin{figure}[!ht]
    \centering
    \includegraphics[width=0.8\linewidth]{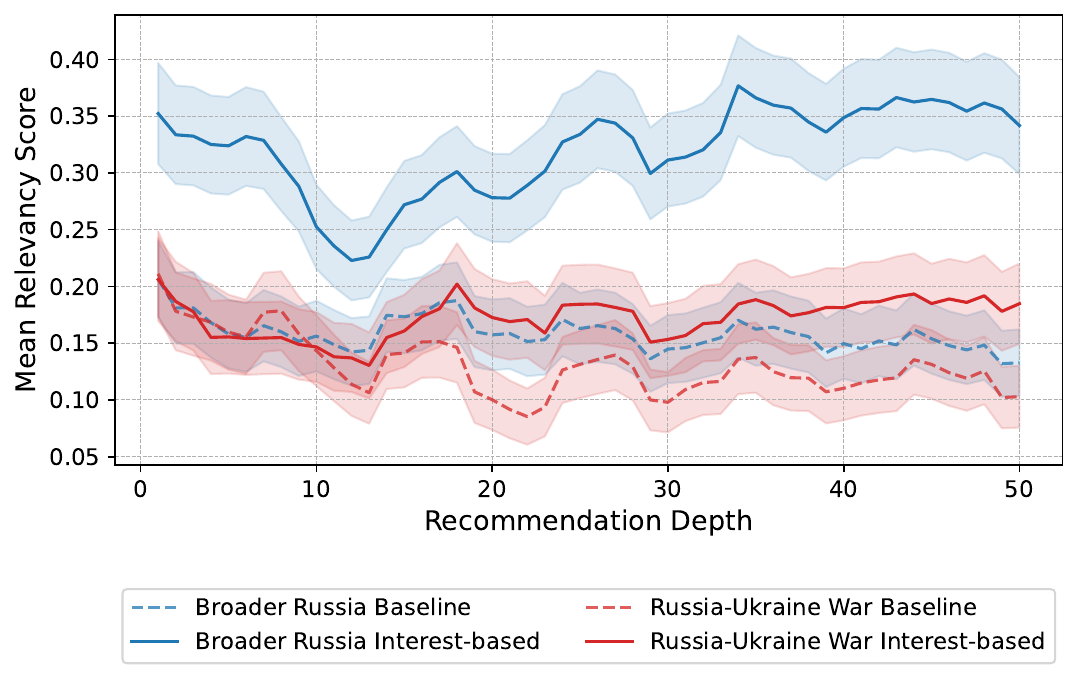}
    \caption{Mean recommendation relevance across depths for the Russia-related topics. The graph compares baseline (3-second skip) and interest-based watch-time conditions for both broader (\textit{Broader Russia}) and more specific (\textit{Russia--Ukraine War}) themes. Shaded regions indicate 95\% confidence intervals.}
    \label{fig:russia_topics}
\end{figure}

\begin{figure}[!ht]
    \centering
    \includegraphics[width=0.8\linewidth]{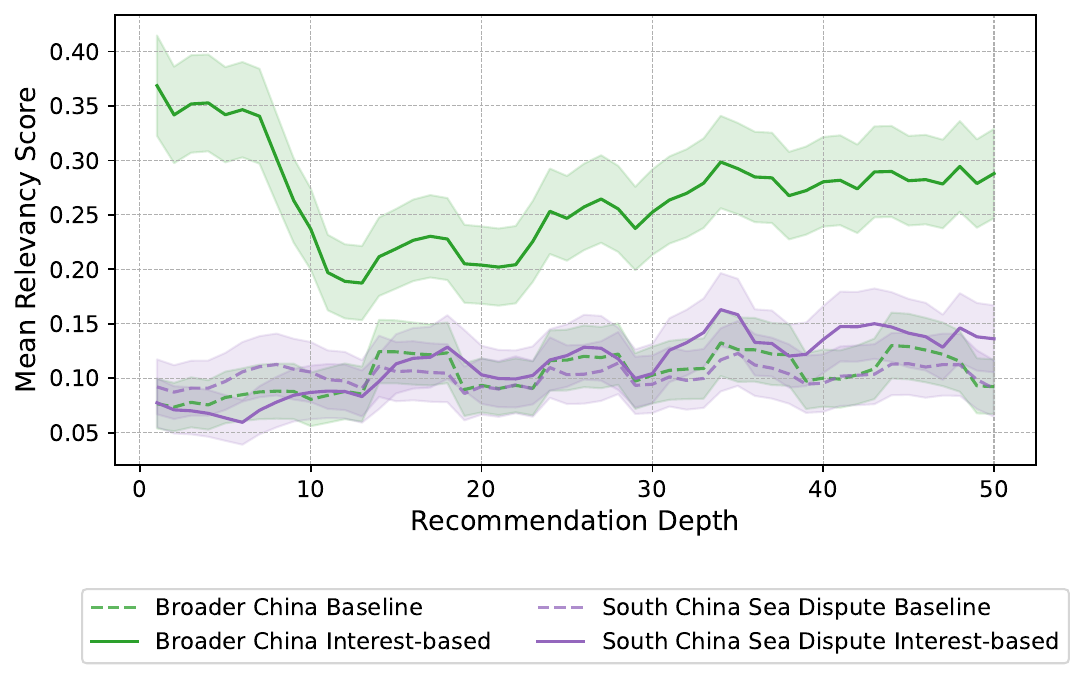}
    \caption{Mean recommendation relevance across depths for the China-related topics. The graph compares baseline (3-second skip) and interest-based watch-time conditions for both broader (\textit{Broader China}) and more specific (\textit{South China Sea Dispute}) themes. Shaded regions indicate 95\% confidence intervals.}
    \label{fig:china_topics}
\end{figure}

Mean relevance scores were compared across 50 depths in the recommendation chain for four thematic datasets: \textit{Broader Russia}, \textit{Russia--Ukraine War}, \textit{Broader China}, and \textit{South China Sea Dispute}. In both Figure~\ref{fig:russia_topics} (Russia topics) and Figure~\ref{fig:china_topics} (China topics), the interest-based watch-time condition consistently produced higher relevance scores across depths compared to the 3-second baseline. These differences emerged early in the chain and persisted through to depth 50.

To quantify the overall recommendation quality, we calculated the area under the curve (AUC) for each recommendation path and performed paired t-tests to evaluate differences between conditions. The AUC represents the cumulative relevance that a user encounters across the entire recommendation trail. Results are summarized in Table~\ref{tab:auc_stats}. For example, Broader Russia’s AUC increased from 8.10 in the baseline condition to 15.92 in the interest-based condition ($t = 18.404$, $p = 2.42 \times 10^{-46}$). Broader China showed a similarly large shift (5.14 to 12.88, $t = 14.387$, $p = 8.79 \times 10^{-30}$). Smaller but statistically significant gains were observed for the Russia--Ukraine War ($\Delta$AUC = 1.63, $p = 1.30 \times 10^{-3}$) and the South China Sea Dispute ($\Delta$AUC = 0.69, $p = 2.79 \times 10^{-2}$).

\begin{table}[!ht]
    \centering
    \caption{AUC comparison between baseline and interest-based watch-time conditions. Paired t-tests compare relevance area under the curve (AUC) across all recommendation depths for each topic.}
    \label{tab:auc_stats}
    \begin{tabular}{lcccc}
        \toprule
        \textbf{Topic} & \textbf{Baseline AUC} & \textbf{Interest AUC} & \textbf{$\Delta$ AUC} & \textbf{p-value} \\
        \midrule
        Broader Russia & 8.10 & 15.92 & 7.82 & $2.42 \times 10^{-46}$ \\
        Russia--Ukraine & 6.82 & 8.45 & 1.63 & $1.30 \times 10^{-3}$ \\
        Broader China & 5.14 & 12.88 & 7.73 & $8.79 \times 10^{-30}$ \\
        South China Sea & 5.06 & 5.75 & 0.69 & $2.79 \times 10^{-2}$ \\
        \bottomrule
    \end{tabular}
\end{table}

While some increase in relevance is expected when a user watches a video for longer, our findings suggest that even small differences in watch time, ranging from 3 to 60 seconds, produce large and persistent shifts in recommendation paths. On a platform like YouTube Shorts, where all videos are under a minute, these differences are minimal in absolute terms but are treated by the system as strong indicators of preference. The result is not merely improved matching, but a rapid narrowing of content diversity and a strong reinforcement of early signals.

This amplification effect is especially evident when comparing broader and more specific topics. Broader themes such as \textit{Russia} and \textit{China} maintained higher relevance throughout the chain, while specific issues like the \textit{Russia--Ukraine War} and \textit{South China Sea Dispute} showed faster drift, particularly in the baseline condition. Even under interest-based viewing, the improvement for narrower topics was much smaller. For example, the South China Sea topic saw only a 0.69 gain in AUC compared to Broader China’s 7.73. This indicates that the recommendation algorithm is less responsive to simulated interest for detailed or complex topics.

Moreover, the overall relevance scores remained low. Across all datasets and conditions, average relevance rarely exceeded 0.35 on a 0--2 scale, suggesting that even with simulated interest, the system delivers a substantial volume of marginally related content. Together, these findings point to a form of algorithmic bias rooted in feedback amplification and topic generalization. Minimal behavioral signals are treated as high-confidence preferences, reinforcing certain content paths while allowing others to fade. In short-form environments where interactions are brief and often ambiguous, this dynamic can limit content diversity and reduce exposure to complex or underrepresented narratives. While sessions were isolated and watch-time was programmatically controlled, external confounders such as trending videos, time of day, or regional content popularity could influence recommendation paths. These variables were not explicitly controlled and represent a source of variation in recommendation behavior.

\section{Conclusion and Discussion}

This study examined how simulated watch-time behavior affects topical relevance in YouTube Shorts recommendations. Using four geopolitical themes and two viewing conditions, uniform short views and interest-based durations, we analyzed how small differences in watch time shape recommendation paths. Relevance was scored using a generative AI model and quantified via AUC and statistical testing.

Our findings directly address the research questions. \textbf{RQ1} is supported by the clear relationship between watch-time and rising relevance across recommendation chains. \textbf{RQ2} is answered by the amplification effect observed under interest-based viewing. \textbf{RQ3} is reflected in the system’s stronger topical retention for broad themes (e.g., \textit{Russia}, \textit{China}) compared to narrower ones (e.g., \textit{Russia--Ukraine War}, \textit{South China Sea Dispute}).

These results highlight a feedback amplification dynamic, where minimal behavioral cues yield strong personalization, reducing content diversity. While our study focused on geopolitical content, the simulation framework can generalize to other domains. Future work will examine additional signals and broader topic categories.

\bmhead{Acknowledgements}

\small

This research is funded in part by the U.S. National Science Foundation (OIA-1946391, OIA-1920920), U.S. Office of the Under Secretary of Defense for Research and Engineering (FA9550-22-1-0332), U.S. Army Research Office (W911NF-23-1-0011, W911NF-24-1-0078), U.S. Office of Naval Research (N00014-21-1-2121, N00014-21-1-2765, N00014-22-1-2318), U.S. Air Force Research Laboratory, U.S. Defense Advanced Research Projects Agency, the Australian Department of Defense Strategic Policy Grants Program, Arkansas Research Alliance, the Jerry L. Maulden/Entergy Endowment, and the Donaghey Foundation at the University of Arkansas at Little Rock. Any opinions, findings, and conclusions or recommendations expressed in this material are those of the authors and do not necessarily reflect the views of the funding organizations. The researchers gratefully acknowledge the support.

\renewcommand{\bibfont}{\fontsize{8.5}{10.4}\selectfont}

\bibliography{sn-bibliography}



\end{document}